\documentclass[9pt,twocolumn,twoside]{opticajnl}
\journal{opticajournal}
\setboolean{shortarticle}{true}

\usepackage{lineno}
\usepackage{siunitx}
\usepackage{ulem}
\usepackage{pdfpages}

\DeclareSIUnit\bar{bar}

\title{Few-cycle optical vortices for strong field physics}

\author[1]{Matthieu Guer}
\author[1]{Martin Luttmann}
\author[1]{Jean-François Hergott}
\author[1]{Fabien Lepetit}
\author[1]{Olivier Tcherbakoff}
\author[1]{Thierry Ruchon}
\author[1,2*]{Romain Géneaux}

\affil[1]{
Universit\'e Paris-Saclay, CEA, LIDYL, 91191 Gif-sur-Yvette, France
}
\affil[2]{
CY Cergy Paris Universit\'e, CEA, LIDYL, 91191 Gif-sur-Yvette, France
}

\affil[*]{romain.geneaux@cea.fr}

\begin{abstract}
We report on the generation of optical vortices with few-cycle pulse durations, 500$\mu$J per pulse, at a repetition rate of 1\,kHz. To do so, a 25\,fs laser beam at 800\,nm is shaped with a helical phase and coupled into a hollow core fiber filled with argon gas, in which it undergoes self phase modulation. 5.5\,fs long pulses are measured at the output of the fiber using a dispersion-scan setup. To retrieve the spectrally resolved spatial profile and orbital angular momentum (OAM) content of the pulse, we introduce a method based on spatially-resolved Fourier transform spectroscopy. We find that the input OAM is transferred to all frequency components of the post-compressed pulse. The combination of these two information shows that we obtain few-cycle, high intensity vortex beams with a well-defined OAM and sufficient energy to drive strong field processes.
\end{abstract}

\setboolean{displaycopyright}{false} 

\begin{document}

\maketitle

\section{Introduction}
Since 1992 and the work of Allen \textit{et al.} \cite{allen_orbital_1992}, it was recognized that orbital angular momentum (OAM) can be imparted to light beams, with ever more pratical implementations~\cite{chen_orbital_2020}. Contrary to the spin angular momentum (SAM), which is related to the local polarization state of the field, a non-zero OAM value is typically associated to an helical wavefront \cite{forbes_structured_2021}. As such they were early refered to as "optical vortices" \cite{coullet_optical_1989}. Prototypical examples of modes with OAM are the Laguerre-Gaussian and Bessel-Gaussian modes \cite{forbes_structured_2021}, which carry a well-defined OAM: all individual photons in such a mode carry the same quantum of OAM \cite{andrews_angular_2013}. This value is given by an azimutal index $\ell$, which corresponds to the number of $2\pi$ phase cycles along a circle centered on the beam axis. The phase singularity in the center of these beams imposes a doughnut-shaped transverse intensity. 

The discovery of vortex beams was followed by numerous applications \cite{padgett_orbital_2017, rosales-guzman_review_2018}
, such as particle manipulation with optical tweezers \cite{he_direct_1995}, stimulated-emission depletion (STED) microscopy \cite{tian_resolution_2015} and optical communications \cite{gibson_free-space_2004, wang_terabit_2012, brunet_design_2014}
. Vortex beams have also been drawing a keen interest in the field of ultrafast light-matter interaction. When brought to femtosecond duration and with sufficient high peak power (above \SI{e14}{\watt \per \centi\meter^2}), they can be used to drive nonlinear optical effects in which the OAM of light becomes a novel quantity to be exchanged in the process. A notable example is the demonstration of extreme ultra-violet (XUV) vortex beams, obtained by high-harmonic generation (HHG) of an infrared vortex driver \cite{hernandez-garcia_attosecond_2013, gariepy_creating_2014, geneaux_synthesis_2016}. Such schemes allow the production of attosecond pulse trains carrying OAM, as well as new forms of structured light \cite{rego_generation_2019, pandey_characterization_2022, luttmann_nonlinear_2023}.

However, optical vortices have yet to be applied in one peculiar field of light-matter interaction: the few-cycle strong field regime, where the duration of light pulses is of the order of magnitude of the optical cycle. The method of post-compression to reach this regime with regular beams is becoming increasingly standard. In a nonlinear medium, the "long" pulse undergoes self-phase modulation (SPM) \cite{stolen_self-phase-modulation_1978, khazanov_post-compression_2022}, which broadens its spectra and thus reduces its Fourier-transform limited duration. To guarantee spatially homogeneous broadening and well-controlled SPM, this process can be realized inside gas-filled hollow-core fibers (HCF) \cite{nisoli_toward_1998, jeong_direct_2018, klas_generation_2020}, multi-pass cells \cite{lavenu_nonlinear_2018, daniault_single-stage_2021, tsai_nonlinear_2022}, or multiple bulk elements \cite{lu_generation_2014, fourmaux_laser_2022}. Such short pulses, when used to drive the HHG process, allow the generation of XUV frequency supercontinua and isolated attosecond pulses \cite{chini_generation_2014, klas_generation_2020}.

\begin{figure*}[t]
    \centering
    \includegraphics[width=\linewidth]{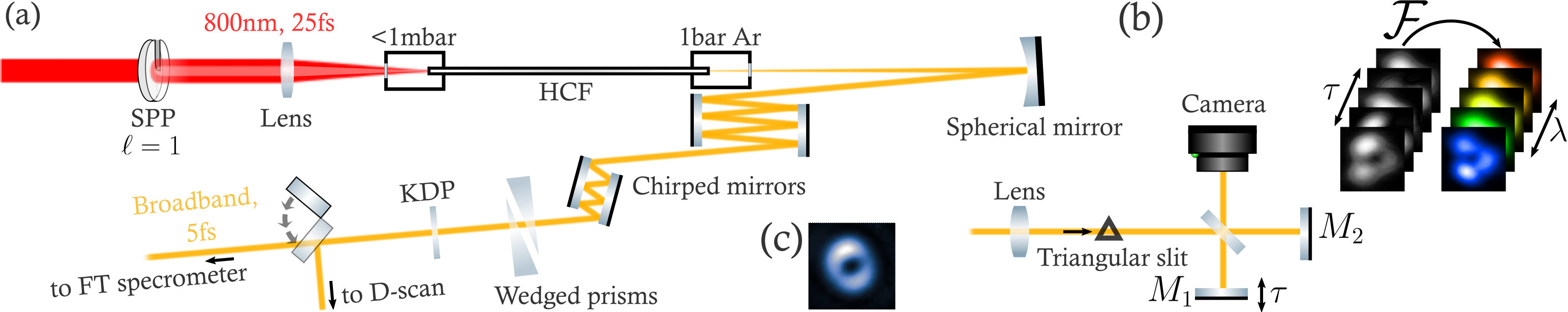}
    \caption{
    (a) Experimental setup. A spiral phase plate (SPP) placed before a hollow-core-fiber (HCF) leads to the generation of a broadband vortex beam.
    (b) Michelson interferometer for the Fourier-transform spectroscopy. An equilateral triangular slit can be inserted before the interferometer, creating a diffraction pattern dependent on the OAM. A series of images is recorded while scaning the delay, and its Fourier transform is computed numerically. 
    (c) Image of the focus after compression.
    }
    \label{fig:setup}
\end{figure*}

In themselves, few-cycle optical vortices (FCOV) are intriguing objects. For instance, there exists a relationship between the value of OAM that a pulse carries and the minimum duration it can have \cite{porras_upper_2019}. Furthermore, using them in nonlinear optics could bring a new field of applications. Such beams with defined OAM, extremely broadband and continuous spectra would make formidable tools to probe recently discovered helical dichroisms, for instance in magnetic structures \cite{fanciulli_observation_2022, mccarter_antiferromagnetic_2023} or in chiral disordered media \cite{rouxel_hard_2022}.

The generation of intense single-cycle vortex beams is technically challenging. Monochromatic vortex beams are typically obtained by converting a Gaussian beam using elements such as spiral phase plates (SPP) \cite{sueda_laguerre-gaussian_2004, beijersbergen_helical-wavefront_1994}, or spatial light modulators \cite{liao_analysis_2013}. These elements imprint an azimutally varying phase on the beam, which corresponds to a well-defined OAM at one specific frequency. 
With single material plates, it was shown in Ref. \cite{miranda_spatiotemporal_2017}, in the low energy regime, that although the central singularity of ultrashort vortex beams is conserved, its OAM content is highly inhomogeneous. Indeed the curling phase imparted ranged from $1.7\pi$ to $2.5\pi$  respectively for 631\,nm and 952\,nm spectral components. It significantly altered the homogeneity of the dounut shaped resulting beam, and would be further amplified in a strongly non linear regime \cite{pandey_characterization_2022}. Other types of devices \cite{beijersbergen_astigmatic_1993} suffer the same kind of limitations in the broadband regime. To circumvent this drawback, achromatic devices have been proposed \cite{xie_optical_2008, naik_ultrashort_2017}, and demonstrated \cite{yamane_ultrashort_2012, grunwald_spatio-temporal_2014}. However, this increases the complexity of the setup and requires potentially expensive material and on demand designs.
Alternatively, theoretical works \cite{cao_relativistic_2020, xu_few-cycle_2022} have recently suggested to first generate a reasonably narrow band optical vortex, and then post-compress it. As such, it was experimentally demonstrated that using SPM in a multi-pass cell for broadening an optical vortex was feasable, and scalable to large power \cite{kaumanns_spectral_2021}, even though the few-cycle regime was not achieved yet in this configuration.

In this letter, we propose and demonstrate a scheme allowing the generation of high intensity few-cycle optical vortices (FCOV) in the visible/near-infrared domain. Instead of trying to convert a few-cycle Gaussian beam into a LG mode, we start from a \SI{25}{\femto\second} Laguerre-Gaussian mode and attempt to post-compress it in an argon-filled stretched hollow-core fiber (HCF).
We show that the vortex beam propagates through the HCF and that its topological charge $\ell$ is not altered by the spectral broadening process. After temporal compression, we characterize the FCOV in space and time, showing that all created frequency components have an OAM of $\ell$ and that the pulse has a duration of two optical cycles.

\section{Experiment}

The sketch of the experimental setup is shown Fig.\,\ref{fig:setup}(a).
A specific in-house amplifier has been developped on the FAB laser~\cite{golinelli_cep-stabilized_2019} of the ATTOLAB facility, delivering \SI{25}{\femto\second},  \SI{2}{\milli\joule} pulses centered at \SI{800}{\nano\metre} at a repetition rate of \SI{1}{\kilo\hertz}, with possible carrier-enveloppe phase (CEP) stabilization.
Prior to post-compression, the Gaussian mode of the beam is passed through a $\ell=1$ SPP, to be imparted OAM.
It is then focused onto the HCF tip.
We use a \SI{1.4}{\metre} long, \SI{400}{\micro \metre} inner diameter HCF filled with argon (few-cycle Inc.). A longitudinal gradient of pressure is applied in order to limit self-focusing and filamentation effects at the entrance of the fiber \cite{suda_generation_2005}. The pressure at the output tip of the fiber is around \SI{1}{\bar}.

Laguerre-Gaussian modes are intuitive to describe OAM-carrying beams in the free-space propagation, however they are not eigenmodes of the fiber.
The modes of a hollow-core fiber are well-documented \cite{degnan_waveguide_1973, heuer_stimulated_1988} and two particular modes, OAM$_{11}^\pm$ of opposite circular polarization \cite{brunet_design_2014}
, display a helical phase similar to the one of the input $\ell=1$ LG beam. 
The laser beam waist at focus is optimized to maximize the overlap with these two modes, similarly to what is usually done with a Gaussian beam \cite{nisoli_toward_1998}. When using a gaussian beam with our fiber, we need a beam waist of \SI{130}{\micro\meter}, however for the Laguerre-Gaussian, we need to adjust it to \SI{110}{\micro\meter}, or 0.56 times the fiber radius (see section 1 of the Supplemental Document). This leads to 95\% coupling between the incident $LG_{1,0}$ mode and the fiber modes.

After spectral broadening in the fiber, the pulse is collimated with a spherical mirror, and then undergoes several reflections on a set of chirped mirrors (PC1332, Ultrafast Innovations) that provide negative group delay dispersion (GDD). The overall GDD of the pulse is finely adjusted by means of two wedge prisms. A \SI{1}{mm}-thick KDP compensates the third order dispersion.
We measure a \SI{0.5}{\milli\joule} pulse energy at the output of the compression setup.

\section{Results}

\begin{figure*}[t!]
    \centering
    \includegraphics[width = \linewidth]{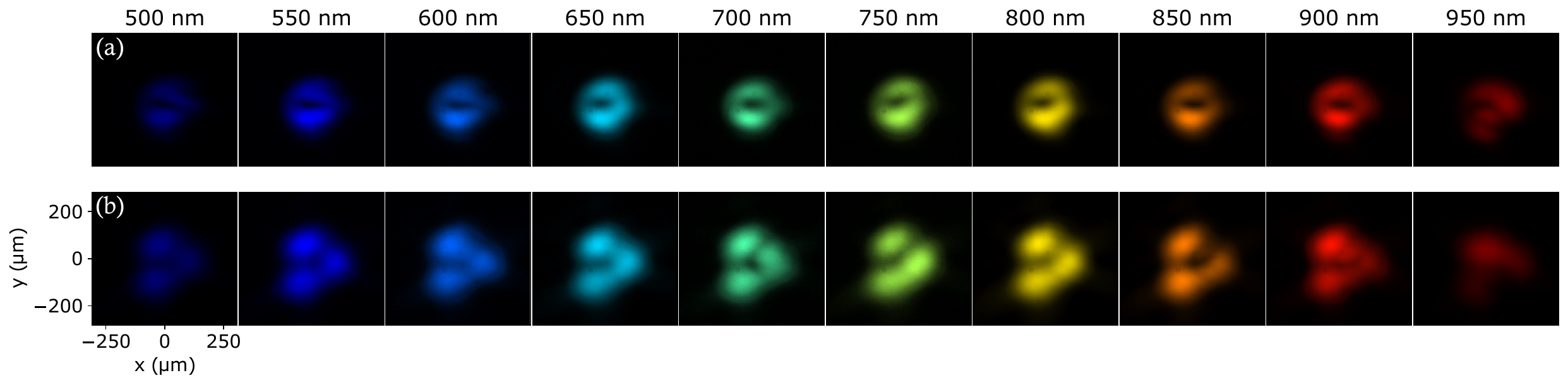}
    \caption{(a) Spectrally resolved transverse intensity of the vortex beam at focus for a selection a wavelengths trhoughout the post-compressed pulse, retrieved by Fourier-Transform spectroscopy. (b) Same when the triangular aperture is inserted at the entrance of the interferometr. For clarity, all images are normalized and displayed in a linear scale going from black to a given color.
    }
    \label{fig:fig2}
\end{figure*}

In order to observe the intensity profile of each frequency component independently, we perform a spatially-resolved Fourier-transform spectroscopy (FTS) experiment \cite{bates_fourier_1978}. The broadband vortex beam is sent into a Michelson interferometer. We record the light interference pattern at the focus of the beam, while the optical path difference between the two arms of the interferometer is varied by steps of \SI{75}{\nano\metre}. All data acquisition is performed using the PyMoDAQ open-source library \cite{weber_pymodaq_2021}. The pixel-wise Fourier transform of the delay-varying intensity provides us with the transverse spatial profile of the beam at each wavelength (Fig.\,\ref{fig:fig2}.a). We observe that each frequency component exhibits a ring shape at focus, typical of light carrying OAM. To confirm the presence of OAM and to measure its value, we insert an equilateral triangular aperture prior to the interferometer. The diffraction pattern by such an aperture is a triangle consisting of several dots, the number of dots on any side of the triangle being equal to $\ell+1$ \cite{hickmann_unveiling_2010}. In our experiment (Fig.\,\ref{fig:fig2}.b), we observe triangles with two dots on each side, corresponding to an OAM quantum $\ell=1$. The pattern is the same at every wavelength, demonstrating that the OAM is transfered to the new frequencies created during self-phase modulation.

To explain the transfer of an equal quantum of OAM to all spectral components, we recall that SPM can be viewed as a degenerate four-wave mixing process, in which two pump photons at $\omega$ carrying $\ell_0=1$ will separate into one Stokes photon at $\omega_s = \omega-d\omega$ and one Anti-Stokes photon at $\omega_{as} = \omega+d\omega$. Because the medium is homogeneous, there is no transfer of OAM to the gas, and the two outgoing photons must carry a total OAM quantum of $\ell_s + \ell_{as}=2\ell_0=2$. The trivial combination is $\ell_s=\ell_{as} =1$, but there could also be other possibilities that satisfy the OAM conservation, for instance $\ell_s=0$ and $\ell_{as} =2$.
However, as demonstrated though the computation of mode overlaps in section 3 of the Supplemental Document, phase matching over the length of the interaction medium effectively suppresses any pathway involving $\ell_{s/as}\neq1$.

\begin{figure}[b!]
    \centering
    \includegraphics[width = 0.85\linewidth]{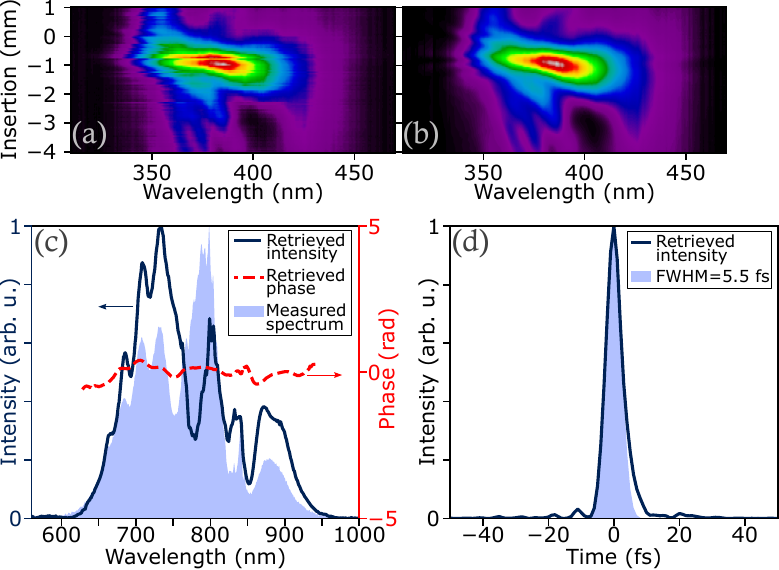}
    \caption{
    (color online) (a) Measured and (b) reconstructed dispersion scans.
    (c) Retrieved spectrum and phase.
    (d) Reconstructed temporal profile, with a FWHM of \SI{5.5}{\femto\second} i.e. 2 optical cycles.
    }
    \label{fig:fig3}
\end{figure}

To complete the characterization of the FCOV, we measure its temporal intensity and phase. To do so, we use the dispersion scan (d-scan) technique \cite{miranda_simultaneous_2012}, which is not perturbed by the OAM carried by the pulse. 
The beam is focused in a \SI{5}{\micro\metre}-thick BBO crystal optimized for second harmonic generation (SHG). A glass wedge placed at the Brewster angle separates the generated second harmonic from the fundamental. We record the second harmonic spectrum while scanning the dispersion in a controlled way, here through the translation of a wedge prism (see Fig.\,1). The spectrum is measured using an integrating sphere, meaning that the result is the spatially-averaged temporal profile. A retrieval algorithm then reconstructs the pulse in amplitude and phase by minimizing the error between the measured d-scan trace and a simulated trace. We use the common pulse retrieval algorithm (COPRA) \cite{geib_common_2019}, as implemented in the open-source retrieval software PyMoDAQ-Femto \cite{weber_femtosecond_2022}. The result of the procedure is presented in Fig.\,\ref{fig:fig3}.  The fundamental and measured spectra have similar shapes but do not match perfectly. This is a known effect in dispersion scans in the presence of experimental noise \cite{miranda_fast_2017}. Despite this discrepancy, the error between the retrieved and experimental traces, using the normalized root mean square error \cite{geib_common_2019}, is \num{7.4e-3} on a 101 x 2048 grid, which guarantees that the temporal profile is retrieved accurately.
 The pulse has a FWHM duration of \SI{5.5}{\femto\second} (Fig.\,\ref{fig:fig3}.d) at the optimal glass insertion. We specify that the CEP was not stabilized during the measurement. For LG modes, CEP variations translate into a random azimutal rotation of the pulse inside its envelope, which has no effect in the d-scan measurement.

The output power of the system is \SI{0.5}{\milli\joule}, which amounts to a rather poor transmission of 25\%.
For comparison, with the same setup but using a Gaussian input beam, with some changes to accomodate the higher peak intensity and different mode size (see section 2 of Supplemental Document), we obtain 50\% transmission and \SI{4.9}{\femto\second} output pulse duration. There is thus a significant drop of transmission when using the LG beam, which cannot be explained by the attenuation coefficient of the OAM$_{11}^\pm$ fiber modes. Instead, 
we attribute this lower efficiency with LG modes to two effects. First, even if the spiral phase plate gives an OAM $\ell=1$ to the beam, it does not transform the Gaussian mode into a pure LG$_{10}$ mode: 21.5\% of the energy is sent towards higher radial modes, which have worse coupling in the fiber \cite{beijersbergen_helical-wavefront_1994}.
Second, the propagation of OAM modes is more sensitive to imperfections in the wavefront than the fundamental mode. This means that all detrimental experimental effects, such as imperfect mode quality of the laser, unwanted nonlinearities, or large B-integrals, will lead to increased loss for LG modes. Notably, we noticed that, as currently implemented, the fiber input window causes measurable nonlinearities. Furthermore, report of higher energy transmission of vortex in HCF~\cite{lin_intense_2018}, although in less demanding conditions, leads us to believe that the performances of our setup could be improved by a better engineered post-compression system and by a more optimized LG conversion technique. 

We can also see in Fig.\,\ref{fig:fig3}.c that the spectrum shows modulations. This indicates a significant level of ionization in the fiber. Indeed, we are using argon, which has a low ionization potential. This choice was motivated because of the lower energy density of the Laguerre-Gaussian mode compared to a Gaussian mode, which means we need a gas with a higher nonlinear index to broaden the spectrum significantly. Thus, we have to compromize between either a larger spectrum (and shorter pulses) or a cleaner one. 
A possible improvement would be to use a fiber with a smaller diameter in order to have a higher energy density in the fiber, and be able to use a gas with lower nonlinear index but a higher ionization potential, such as neon or helium.
Nonetheless, given the pulse duration of \SI{5.5}{\femto\second}, and assuming the focal length of \SI{1}{\meter} we used to image the focus in our Michelson interferometer (Fig.\,\ref{fig:fig2}), the irradiance at focus in the current conditions amounts to \SI{1.5e14}{\watt/\centi\meter^2}, which is already sufficient for usage in nonlinear processes such as high-harmonic generation.

\section{Conclusion}
We demonstrated that a \SI{2}{\milli\joule} beam with OAM can be compressed in a hollow-core fiber with about 25\% transmission, leading to a transmitted energy of \SI{0.5}{\milli\joule}. We showed that after the fiber, the OAM content of the beam is dominated by $\ell=1$ at all frequencies.
The resulting pulse duration is 2.1 optical cycles (\SI{5.5}{\femto\second}). Combined together, this means we can already reach values of irrandiance that are enough to drive highly nonlinear processes, such as HHG. 
The process is also scalable to larger energies, through an adaptation of the fiber length, diameter and gas. It is also scalable to larger topological charges, as the fundamental limit of Ref. \cite{porras_upper_2019} is not reached in this configuration. Generating harmonics with this kind of driver would open the path for time-dependent applications of the OAM at the attosecond timescale, and on a very broad continuous spectral range, which is out of reach today.

\begin{backmatter}

\bmsection{Acknowledgments} The authors would like to thank Thierry Auguste and Marc Hanna for fruitful discussions.

\bmsection{Funding} This work was supported by the French Agence Nationale pour la Recherche (under grants TOCYDYS, ANR-19-CE30-0015-01 and HELIMAG, ANR-21-CE30-0037), the European Union (ERC, Spinfield, Project No. 101041074 and Horizon 2020 Programme No. EU-H2020-LASERLAB-EUROPE-654148), and Investissements d'Avenir of LabEx PALM (ANR-10-LABX-0039-PALM).

\bmsection{Disclosures} The authors declare no conflicts of interest.

\bmsection{Data availability} Data underlying the results presented in this paper are available from the authors upon reasonable request.

\bmsection{Supplemental document}
See Supplement 1 for supporting content. 

\end{backmatter}

\bigskip

\bibliography{Biblio_zot}
\bibliographyfullrefs{Biblio_zot}

\clearpage


\section*{Supplementary material}

# Few-cycle optical vortices for strong field physics: supplemental document

## 1. COUPLING INTO THE FIBER

The coupling efficiency can be estimated by [1]:

$$\eta = \frac{(\int E_1 E_2^*)^2}{\int |E_1|^2 \int |E_2|^2}, \tag{S1}$$

where $\eta$ is the coupling efficiency, and $E_1$ and $E_2$ are the transverse profile of respectively the input beam and the fiber mode. The integral is carried out over transverse dimensions. The fiber modes carrying OAM are described in [2–4]. They are indexed by an azimuthal index $\ell$, equal to the OAM quantum, and a radial index $p$, with $p \geq 1$

We first look at the coupling efficiency between a Gaussian input beam and the fiber modes. The results are shown in Fig. S1(a), and are agreeing with the litterature [1]: the beam couples mainly to the lowest radial mode, with a maximal efficiency of 0.98, for a waist/radius ratio $\frac{W}{a} = 0.66$. No energy is sent towards modes carrying a non-zero OAM. In the case of a Laguerre-Gaussian input (Fig. S1(b)), the result is very similar: the optimal coupling is with the mode $\ell = 1, p = 1$ of the fiber. The maximum of coupling should be 0.96, achieved at $\frac{W}{a} = 0.55$.

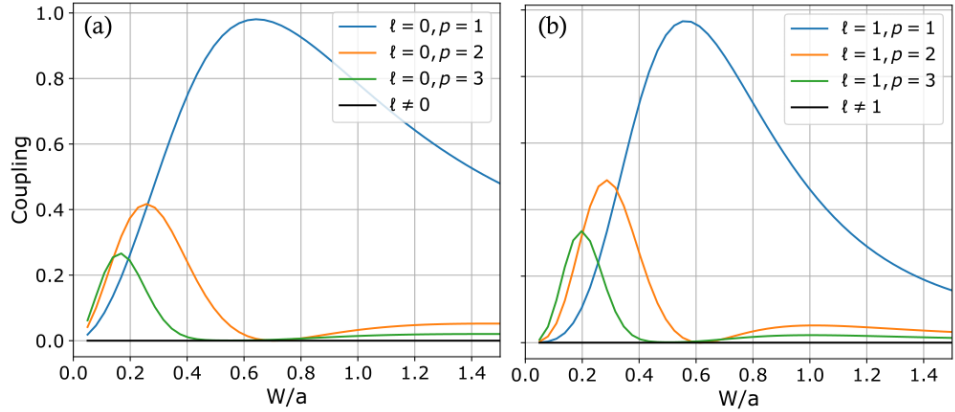

**Fig. S1.** Coupling to the fiber modes for a (a) Gaussian or a (b) Laguerre-Gaussian input, depending on the waist (normalized by the fiber radius). The solid black curve shows the sum of the coupling towards all modes carrying an OAM different from the input beam.

## 2. COMPRESSION OF A GAUSSIAN BEAM

The setup for compression without OAM was a bit different from the one described in the article. To couple in the fiber, we used a Gaussian mode with a waist of 130 µm, for better overlap with the fiber modes carrying no OAM. Since the peak intensity is higher with a Gaussian beam, we use helium instead of argon for our nonlinear medium. This reduces the losses due to ionization, which enables us to get a much higher transmission: 80% without gas, 50% with 4 bar helium. We measured the duration of the pulses with the same method as described in the article, and show the results in figure S2. We find a pulse duration of 4.9 fs, less than two optical cycles.

## 3. ORBITAL ANGULAR MOMENTUM IN SPM

The probability amplitude to generate a specific combination of Stokes and Anti-Stokes modes during the self-phase modulation process is given by their overlap with the pump profiles [5]:

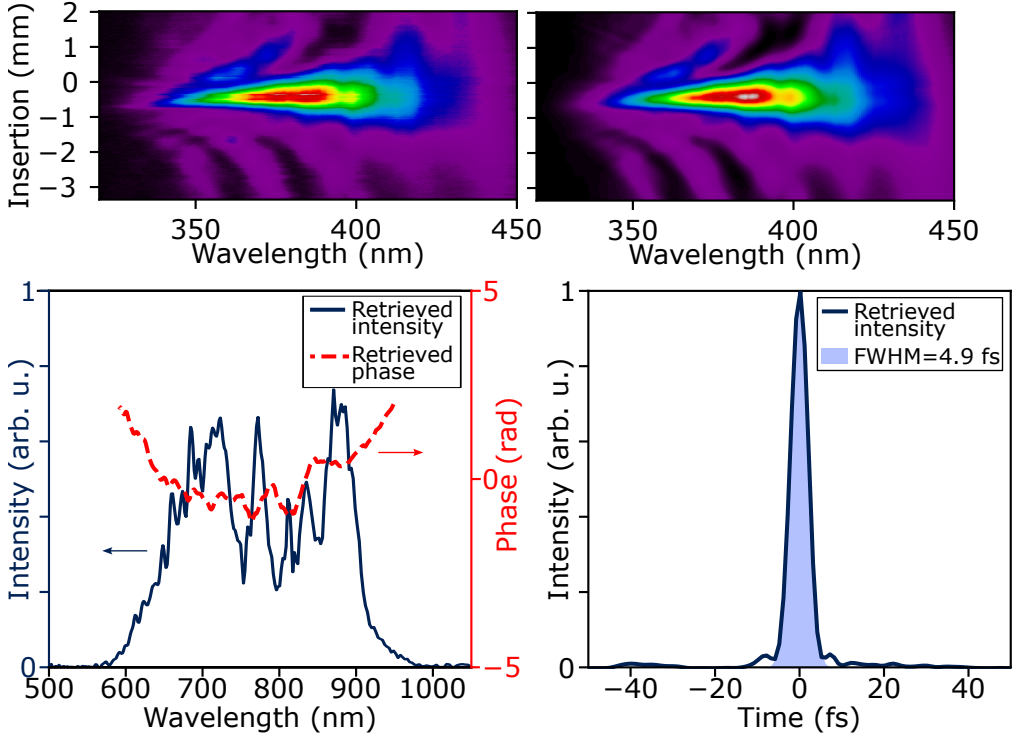

**Fig. S2.** Dispersion scan trace measured an retrieved, retrieved spectrum and pulse, same as Fig. 3 of the main text, but for a Gaussian input pulse.

$$c_{\ell_0,\ell_s,\ell_{as}} = \int \left(u_{\ell_0}^{\omega_0}\left(\mathbf{r}\right)\right)^2 \left(u_{\ell_s}^{\omega_s}\left(\mathbf{r}\right) u_{\ell_{as}}^{\omega_{as}}\left(\mathbf{r}\right)\right)^* d\mathbf{r} \tag{S2}$$

where $u_{\ell_x}^{\omega_x}$ is the field of a Laguerre-Gaussian mode with azimuthal index $\ell_x$ and waist $\omega_x$. We numerically evaluate this integral over the transverse plane for several pairs of $(\ell_s, \ell_{as})$. We consider a Stoke shift of 5 nm, and apply the Boyd criterion [5]: the waists of the generated fields are set so that their Rayleigh ranges match that of the pump fields. Figure S3(b) shows that the overlap is much larger for the $\ell_s = \ell_{as} = 1$ case. Moreover, the same situation takes place when integrating along the fiber propagation dimension, strenghtening the dominance of this pathway over the other quantum paths.

This is even further amplified by phase matching effects: in the fiber, the different modes have different propagation constants (even without gas), whose formula can be found in [6]. We computed the phase shift between the $\ell = 1$ driver mode, and several other modes, and represented it in figure S3(c). We can see that the phase mismatch accumulated on the whole fiber between two different values of OAM far exceeds $2\pi$, meaning that any mode with $|\ell| \neq 1$ is suppressed by phase-matching. Thus, even though the SPM process by itself does not transfer its OAM content of the driver to the new frequencies, our long interaction medium filters out values of OAM different from the driver's.

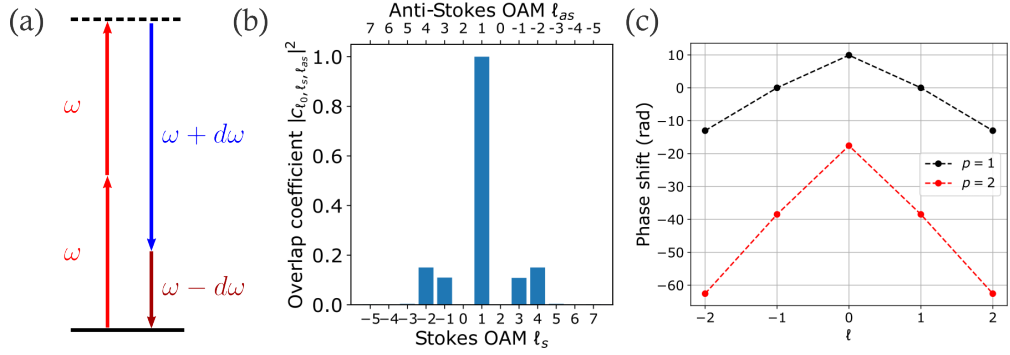

**Fig. S3.** (a) Self-phase modulation as a four photons process. Two photons from the driver at $\omega$ separate one Stokes photon at $\omega_s = \omega - d\omega$ and one Anti-Stokes photon at $\omega_{as} = \omega + d\omega$. (b) Probability of the SPM yielding a photon pair $(\ell_s, 2\ell_0 - \ell_s)$, given by the overlap of their respective modes (eq. S2). (c) Phase difference accumulated compared to the driver by modes of various $\ell$ and $p$ indexes during the propagation on the fiber length.


\end{document}